\input ./style/arxiv-general.cfg
\documentclass[MSNbibl,nameyear,dvips]{arxstspdf}
\makeatletter
   \@ifpackageloaded{graphicx}{}{\usepackage{graphicx}}
\makeatother
\usepackage{flushend}
\usepackage{stfloats}

\volume{30}
\issue{2}
\pubyear{2015}
\firstpage{164}
\lastpage{166}
\doi{10.1214/15-STS515}
\referstodoi{10.1214/14-STS487}
\docsubty{FLA}

\makeatletter
\newcommand{\eqref}[1]{(\ref{#1})}

\def\mm#1{\mathbf{#1}} 
\def\mmm#1{\bolds{#1}}
\makeatother

\begin{document}
\begin{frontmatter}
\vspace*{12pt}
\title{Beyond the Valley of the Covariance Function}
\runtitle{Comment}

\begin{aug}
\author[A]{\fnms{Daniel}~\snm{Simpson}\ead[label=e1]{D.P.Simpson@warwick.ac.uk}},
\author[B]{\fnms{Finn}~\snm{Lindgren}} 
\and
\author[C]{\fnms{H\aa{}vard}~\snm{Rue}\corref{}\ead[label=e3]{hrue@math.ntnu.no}}
\runauthor{D. Simpson, F. Lindgren and H. Rue}

\affiliation{University of Warwick, University of Bath and Norwegian
University of
Science and Technology}

\address[A]{Daniel Simpson is CRiSM Fellow, Department of Statistics,
University of Warwick, CV4 7AL,
Coventry, United Kingdom \printead{e1}.}
\address[B]{Finn~Lindgren is Reader in Statistics, Department of Mathematical
Sciences,
University of Bath,
BA2 7AY,
Bath, United Kingdom.}
\address[C]{H\aa{}vard Rue is Professor of Statistics, Department of Mathematical Sciences,
Norwegian University of
Science and Technology,
N-7491,
Trondheim, Norway \printead{e3}.}
\end{aug}


\end{frontmatter}

\section{Introduction}

Multivariate models are under-represented in the literature on spatial
statistics. There is a basic reason for this: univariate models are
sufficiently complicated to keep us busy. Genton and Kleiber have
done a fabulous job compiling and investigating the available models,
with a focus on the important class of models that they, with
collaborators, introduced. This paper gives a solid state of the art
and points out just how many holes there are in the theory and
practice associated with these fields. This gives us licence to point
out some other holes and to suggest some important directions for the
future.

\section{There is Power in a Spectrum}

If we were to quibble about one thing in Genton and Kleiber's paper,
it would be that we disagree over the extent to which the class of
multivariate GRFs has been categorized. Note that this is different
from \emph{explicitly} constructing valid cross-covariance functions! To
wit, if a multivariate GRF has a spectral representation, the spectral
representation given in Section~1.2 completely characterizes the class
of stationary multivariate random fields that admit an absolutely
continuous spectral measure. This represents a large chunk of
interesting GRFs. We note that the paper, by restricting the
cross-spectral densities to be real, implicitly assumes that
$C_{ij}(\mm{h})=C_{ij}(-\mm{h})$, when the minimal necessary
requirement is only that $C_{ij}(\mm{h})=C_{ji}(-\mm{h})$, which
allows for phase differences between the model components. The
representation can then be employed constructively as follows. Let
$\mmm{\omega} \rightarrow\mm{S}(\mmm{\omega})$ be a mapping from
$\mathbb{R}^d$ to the set of Hermitian nonnegative definite matrices,
the elements of which the cross-spectral densities, denoted $f_{ij}$
in the paper, are here subject to
$f_{ij}(\mmm{\omega})=\overline{f_{ji}(\mmm{\omega})}$.
Then, for any complex, matrix-valued function $\mm{L}(\mmm{\omega})$
such that
$L_{ij}(\mmm{\omega})=\overline{L_{ij}(-\mmm{\omega})}$ and
$\mm{S}(\cdot) = \mm{L}(\cdot) \overline{\mm{L}(\cdot)}$,
%
\begin{eqnarray}
\label{eq:spect} \mm{x}(\mm{s}) &=& \int_{\mathbb{R}^d} \mm{L}(\mmm{
\omega}) \mathrm{e}^{i \mm{s}\cdot
\mmm{\omega}} \,d\widetilde{\mm{W}}(\mmm{\omega})
\nonumber
\\[-8pt]
\\[-8pt]
\nonumber
& = &\int
_{\mathbb{R}^d}\int_{\mathbb{R}^d} \mm{L}(\mmm{\omega})
\mathrm{e}^{i (\mm{s} - \mm{s}')\cdot
\mmm{\omega}} \,d\mmm{\omega} \,d\mm{W}\bigl(\mm{s}'\bigr),
\end{eqnarray}
where $d\widetilde{\mm{W}}(\cdot)\in\mathbb{C}^p$ and
$d\mm{W}(\cdot)\in\mathbb{R}^p$ are Gaussian white noise processes on
$\mathbb{R}^d$ understood as random measures with
$d\widetilde{{W}}_i(\mmm{\omega})=\overline{d\widetilde{{W}}_i(-\mmm
{\omega})}$,
$\mathbb{E} [
d\widetilde{\mm{W}}(\mmm{\omega})\cdot\break  \overline{d\widetilde{\mm
{W}}(\mmm{\omega}')}
 ]=
\delta(\mmm{\omega}-\mmm{\omega}')\mm{I}\,d\mmm{\omega}$,
and
$\mathbb{E} [
d\mm{W}(\mm{s}) \,\overline{d\mm{W}(\mm{s}')}
 ]=\delta(\mm{s}-\mm{s}')\mm{I}\,d\mm{s}$
(\cite{book104}; \cite{lindgrenbook1}). This representation
only covers multivariate GRFs with absolutely continuous spectral
measures; however, the same procedure applies to fields with an atomic
spectral representation. The abstract feature that is hiding in all of
this specificity is that we are explicitly constructing a square root
of the multivariate covariance operator and using this square root to
filter the multivariate white noise. On a compact domain, the
covariance operator is a compact, trace class operator, and so this
square root is well defined using the usual functional calculus.

Another reason to further emphasize this spectral representation is that
it is not only constructive in its own right, but also useful when
transformed back to the nonspectral domain. Kernel convolution
methods (\cite{art473}) have a storied history in univariate spatial
statistics and their generalization to the multivariate case is
straightforward (\cite{simpson2012order}; \cite{Bolin2011}). Their advantage
is that it is never necessary to identify the spectrum of the process
or, in fact, the cross-covariance structure. Rather, for any $L^2$
matrix-valued function $\mm{K}(\cdot,\cdot)$, $\mm{x}(s) = \int
\mm{K}(\mm{s},\mm{s}') \,d \mm{W}(\mm{s}')$ is a valid mutlivariate
GRF, which can be approximated by (carefully) approximating the
corresponding integral with a sum. In fact, there is nothing special
about white noise in this situation, $\mm{W}(\cdot)$ can be any
independently scattered $\mathbb{R}^d$-valued random measure. This
leads to a natural way to construct non-Gaussian random
fields (\cite{aaberg2011class}).

Different choices of $\mm{K}(\cdot,\cdot)$ generating the same
covariance function will in the non-Gaussian case affect the
dependence structure. Similarly, the choice of square root in the
spectral representation becomes relevant, and the two integrals in
\eqref{eq:spect} are no longer guaranteed to give the same process
model. We also note that $\mm{K}(\mm{s},\mm{s}')$ does not have to be
a function of $\mm{s} - \mm{s}'$ and, hence, this resulting field does
not need to be stationary.

An important advantage to the multivariate spectral and convolution
kernel constructions is that the spectral operator
$\mm{L}(\mmm{\omega})$ and the kernel matrix $\mm{K}(\cdot,\cdot)$ can
reflect the modeler's knowledge about the physical process under
consideration. This can lead to useful, informative covariance
structures that are tailored to the specific application. This idea
falls under the auspices of the physics-constrained cross-covariance
specifications mentioned in Section~7.3, except for one key
difference: \emph{while a cross-covariance structure is present,
identifying it is not necessary for inference, either conceptually
or computationally}!

\section{Think Local, Act Local}
In fact, there are many parts of the above construction that we can
live without. The SPDE approach introduced by \citet{Lindgren2011} as
a computationally efficient reformulation of GRFs is an example of the
same procedure where we never construct the kernel matrix. Instead,
multivariate models can be constructed using systems of equations,
such that the kernel corresponds to (matrix-valued) Green's
function of some linear partial differential
operator~(\citeauthor{tech113}, \citeyear{tech113,tech112,tech114}). Using the partial
differential operator construction has several advantages over the
direct kernel specification. First, when the representation is
Markovian, it allows us to localize the process. This is especially
convenient when moving to non-Euclidean spaces; the great tragedy of
spatial statistics is that the Earth turned out not to be flat. The
second major advantage is that this localization allows us to
construct local approximations to the resulting random fields. For an
$n$-dimensional approximation to a field observed at $N$ points, this
reduces the computational cost of fitting the field from
$\mathcal{O}((pN)^3)$ to $\mathcal{O}(pN + p^3n^{3/2})$, for the case
$d=2$. This also compares well with non-Markovian methods, such as
the aforementioned $n$-dimensional convolution kernel methods that
have computational cost of $\mathcal{O}(pn^2 N +p^3n^3)$. Given that
we are now living in the age of ``big data,'' this is a serious
advantage to the SPDE specification. The third major advantage is
that it is straightforward to construct nonstationary models by
locally varying the partial differential operator in the model. This
corresponds to the ``physics'' view, where dependency structures are
specified locally and extended to a global covariance structure
through a conditioning argument. In our experience, this is an
extremely powerful tool for specifying useful covariance and
cross-covariance structures.

\section{With Low Power Comes Great Responsibility}

One of the principal challenges that we have encountered when applying
likelihood methods to multivariate GRFs is that their likelihood
surfaces tend to be flat. This is perhaps not a surprise. If fitting
$p$ univariate models requires the estimation of $\mathcal{O}(p)$
parameters, then a $p$-component multivariate model will require the
same data to be informative about $\mathcal{O}(p^2)$ parameters. This
parameter inflation becomes noticeable already for bivariate
models, but for large $p$ it is a serious issue, that can sometimes
be partially alleviated by using a low rank model, ideally motivated
by problem-specific knowledge. Another option is to impose a sparse
structure on the linear filter matrix operator or to impose
constraints between the parameters.

While flat likelihoods resulting from an exploding parameter space are
annoying, there is a far more pathological problem for univariate
GRFs. Fundamentally, the range and variance parameters are not
identifiable under the usual infill regime (\cite{Zhang2004}). This
leads to a ridge in the parameter space that can only be resolved
using careful prior modeling (\cite{simpson2014penalising}). Ridges
will also seriously challenge numerical optimizers and MCMC schemes
unless they use enough second-order information to resolve it. It is
currently unclear to what extent these problems extend to multivariate
models; however, we suspect that they do, due to the aforementioned
parameter inflation.

%




\end{document}